\documentclass[toc]{PoS}
\usepackage{graphicx}
\usepackage{amssymb}
\usepackage{fontenc}
\usepackage{times}
\usepackage{mathptmx}

\newcommand{\be}{\begin{equation}}
\newcommand{\ee}{\end{equation}}
\newcommand{\bea}{\begin{eqnarray}}
\newcommand{\eea}{\end{eqnarray}} \newcommand{\nn}{\nonumber}
\newcommand{\dd}{\displaystyle}
\newcommand{\de}{\partial}

\def\slash#1{\setbox0=\hbox{$#1$}#1\hskip-\wd0\dimen0=5pt\advance
       \dimen0 by-\ht0\advance\dimen0 by\dp0\lower0.5\dimen0\hbox
         to\wd0{\hss\sl/\/\hss}}

\title{Color Superconductivity in High Density QCD}
\ShortTitle{Color Superconductivity in High Density QCD}

\author{Roberto Casalbuoni\\
    Dipartimento di Fisica dell' Universita' di Firenze and Sezione
INFN\\ Via G. Sansone 1, 50019 Sesto Fiorentino (Firenze), Italy.
\\
    E-mail: \email{casalbuoni@fi.infn.it}}

\abstract{We describe the effects of the strange quark mass and of
color and electric neutrality on the superconducing phases of QCD.
We discuss various phases pointing out the corresponding problems,
typically arising from a chromomagnetic instability. We show that,
at least for particular configurations, the LOFF phase dominates
over the gCFL phase, when close to the transition from gCFL to the
normal phase.}

\FullConference{29$^{\rm th}$ Johns Hopkins Workshop in Theoretical
Physics: Strong Matter in the Heavens\\
Budapest, Hungary \\
August 1-3, 2005}

\begin{document}

\section{Introduction}

It is now a well established fact that at zero temperature and
sufficiently high densities quark matter is a color superconductor
\cite{barrois,cs} (see also Alford and Rischke contributions at this
workshop). The study starting from first principles was done in
\cite{weak,PR-sp1,weak-cfl}. At chemical potentials much higher than
the masses of the quarks $u$, $d$ and $s$, the favored state is the
so-called Color-Flavor-Locking (CFL) state, whereas at lower values
the strange quark decouples and the relevant phase is called
two-flavor color superconducting (2SC).

An interesting possibility is that in the interior of compact
stellar objects (CSO) some color superconducting phase might exist.
In fact we recall  that the central densities for these stars could
be up to $10^{15}$ g/cm$^{3}$, whereas the temperature is of the
order of tens of keV. However the usual assumptions leading to prove
that for three flavors the favored state is CFL should now be
reviewed. Matter inside a CSO should be electrically neutral and
should not carry any color. Also conditions for $\beta$-equilibrium
should be fulfilled. As far as color is concerned, it is possible to
impose a simpler condition, that is color neutrality, since in
\cite{Amore:2001uf} it has been shown that there is no free energy
cost in projecting color singlet states out of color neutral states.
Furthermore one has to take into account that at the interesting
densities the mass of the strange quark is a relevant parameter. All
these three effects:
\begin{enumerate}
  \item mass of the strange quark,
  \item$\beta$-equilibrium,
  \item color and electric neutrality
\end{enumerate}
imply that the radii of the Fermi spheres of the quarks that would
pair are not of the same size, thus creating a problem with the
usual BCS pairing. Let us start from the first point. Suppose to
have two fermions of masses $m_1=M$ and $m_2=0$ at the same chemical
potential $\mu$. The corresponding Fermi momenta are \be
p_{F_1}=\sqrt{\mu^2-M^2},~~~p_{F_2}=\mu.\ee Therefore the radius of
the Fermi sphere of the massive fermion is smaller than the one of
the massless particle. If we assume $M\ll \mu$  the massive particle
has an effective chemical potential
\be\mu_{\rm{eff}}=\sqrt{\mu^2-M^2}\approx \mu-\frac{M^2}{2\mu},\ee
and the mismatch between the two Fermi spheres is
\be\delta\mu\approx\frac{M^2}{2\mu}.\label{eq:3}\ee This shows that
the quantity $M^2/(2\mu)$ behaves as a chemical potential. Therefore
for $M\ll\mu$  the mass effects can be taken into account through
the introduction of the  mismatch between the chemical potentials of
the two fermions given by eq. (\ref{eq:3}). This is  the way  we
will follow in our study.

Now let us discuss the $\beta$-equilibrium. If electrons are present
(as generally required by electrical neutrality) chemical potentials
of quarks of different electric charge are different. In fact, when
at the equilibrium for the process $d\to ue\bar\nu$ we have \be
\mu_d-\mu_u=\mu_e.\ee From this condition we get  that for a quark
of charge $Q_i$ the chemical potential $\mu_i$ is given by \be
\mu_i=\mu+Q_i\mu_Q,\ee where $\mu_Q$ is the chemical potential
associated to the electric charge. Therefore \be\mu_e=-\mu_Q.\ee
Notice also that $\mu_e$ is not a free parameter since it is
determined by the neutrality condition \be Q=-\frac{\de \Omega}{\de
\mu_e}=0.\ee At the same time the chemical potentials associated to
the color generators $T_3$ and $T_8$ are determined by the color
neutrality conditions \be\frac{\de \Omega}{\de \mu_3}= \frac{\de
\Omega}{\de \mu_8}=0.\label{eq:1}\ee

We see that the general there is a mismatch between the quarks that
should pair according to the BCS mechanism for $\delta\mu=0$.
Therefore, in general, the system will go to a normal phase, since
the mismatch, as we shall see, tends to destroy the BCS pairing, or
a different phase will be formed. In the next Sections we will
explore some of these possible phases.

\section{Pairing Fermions with Different Fermi Momenta}

In order to discuss the pairing of fermions with different Fermi
momenta let us review the gap equation for the BCS condensate. The
condensation phenomenon is the key feature of a degenerate Fermi gas
with attractive interactions. Once one takes into account the
condensation the physics can be described using the Landau's idea of
quasi-particles. In this context quasi-particles are nothing but
fermionic excitations around the Fermi surface described by the
following dispersion relation \be \epsilon(\vec
p,\Delta_0)=\sqrt{\xi^2+\Delta_0^2},\ee with \be \xi=E(\vec
p)-\mu\approx \frac{\partial E(\vec p)}{\partial \vec p}\Big|_{\vec
p={\vec p}_F}\cdot (\vec p-{\vec p}_F)={\vec v}_F\cdot(\vec p-{\vec
p}_F)\label{xi},\ee and $\Delta_0$ the BCS condensate. The
quantities ${\vec v}_F$ and $(\vec p-{\vec p}_F)$ are called the
Fermi velocity and the residual momentum respectively. A easy way to
understand how the concept of quasi-particles comes about in this
context is to study the gap equation at finite temperature. For
simplicity let us consider the case of a four-fermi interaction. The
euclidean gap equation is given by \be
1=g\int\frac{d^4p}{(2\pi)^4}\frac 1{p_4^2+|\vec
p\,|^2+\Delta^2_0}.\ee From this expression it is easy to get the
gap equation at finite temperature. We need only to convert the
integral over $p_4$ into a sum over the Matsubara frequencies \be
1=gT\int\frac{d^3
p}{(2\pi)^3}\sum_{n=-\infty}^{+\infty}\frac{1}{((2n+1)\pi
T)^2+\epsilon^2(\vec p,\Delta_0)}.\ee Performing the sum we get \be
1=\frac{g}2\int\frac{d^3p}{(2\pi)^3}\frac{1-n_u-n_d}{\epsilon(\vec
p,\Delta_0)}.\label{gap_T_finite}\ee Here $n_u$ and $n_d$ are the
finite-temperature distribution functions for the excitations
(quasi-particles) corresponding to the original pairing fermions \be
n_u=n_d=\frac 1{e^{\epsilon(\vec p,\Delta_0)/T}+1}.\ee At zero
temperature ($n_u=n_d\to 0$) we find (restricting the integration to
a shell around the Fermi surface)\be
1=\frac{g}2\int\frac{d\Omega_p\, p_F^2\,
d\xi}{(2\pi)^3}\frac{1}{\sqrt{\xi^2+\Delta_0^2}}.\ee In the limit of
weak coupling we get \be \Delta_0\approx 2\,\bar\xi\,
e^{-2/(g\rho)},\label{BCS}\ee where $\bar\xi$ is a cutoff and \be
\rho=\frac{p_F^2}{\pi^2 v_F}\ee is the density of states at the
Fermi surface. This shows that decreasing the density of the states
the condensate decreases exponentially. From a phenomenological
point of view, one determines the coupling $g$ requiring that the
same four-fermi interaction, at zero temperature and density, gives
rise to a constituent mass of the order of $400~MeV$. From this
requirement, using  values for $\mu\approx 400\div 500~MeV$
(interesting for the physics of compact stellar objects),  one
obtains values of $\Delta_0$ in the range $20\div 100~MeV$. However,
since at very high density it is possible to use perturbative QCD,
one can evaluate the gap from first principles \cite{weak}. The
result is \be \Delta_0\approx 2b\mu  e^{-3\pi^2/\sqrt{2}g_s},\ee
with \be b\approx 256\pi^4\left(2/N_f\right)^{5/2} g_s^{-5}.\ee It
is interesting to notice that from Nambu-Jona Lasinio type of models
one would expect a behavior of the type $\exp(-c/g_s^2)$ rather than
$\exp(-c/g_s)$. This is due to an extra infrared singularity from
the gluon propagator. Although this result is strictly valid only at
extremely high densities, if extrapolated down to densities
corresponding to $\mu\approx 400\div 500~MeV$, one finds again
$\Delta_0\approx 20\div 100~MeV$.

We start now our discussion considering a simple model with two
pairing quarks, $u$ and $d$,  with  chemical potentials
\be\mu_u=\mu+\delta\mu,~~~~\mu_d=\mu-\delta\mu,\ee and no further
constraints. The gap equation has the same formal expression as
given in eq. (\ref{gap_T_finite}) for the BCS case, but now
$n_u\not=n_d$ \be n_{u,d}=\frac 1{e^{(\epsilon(\vec
p,\Delta)\pm\delta\mu)/T}+1}.\ee  In the limit of zero temperature
we obtain \be
1=\frac{g}2\int\frac{d^3p}{(2\pi)^3}\frac{1}{\epsilon(\vec
p,\Delta)}\left(1-\theta(-\epsilon-\delta\mu)-\theta(-\epsilon+\delta\mu)
\right)\label{gap}.\ee The meaning of the two step functions is that
at zero temperature there is no pairing when $\epsilon(\vec
p,\Delta)<|\delta\mu|$. In other words the pairing may happen only
for excitations with positive energy. However, the presence of
negative energy states, as in this case, implies that there must be
gapless modes. When this happens there are blocking regions in the
phase space, that is regions where the pairing cannot occur.  The
effect is to inhibit part of the Fermi surface to the pairing giving
rise a to a smaller condensate with respect to the BCS case where
all the surface is used. In the actual case the gap equation at
$T=0$ has two different solutions (see for instance ref.
\cite{Casalbuoni:2003wh}) \be a)~~~
\Delta=\Delta_0,~~~~~~b)~~~\Delta^2=2\delta\mu\Delta_0-\Delta_0^2,\ee
\begin{figure}[h]
\begin{center}
  \includegraphics[width=.5\textwidth]{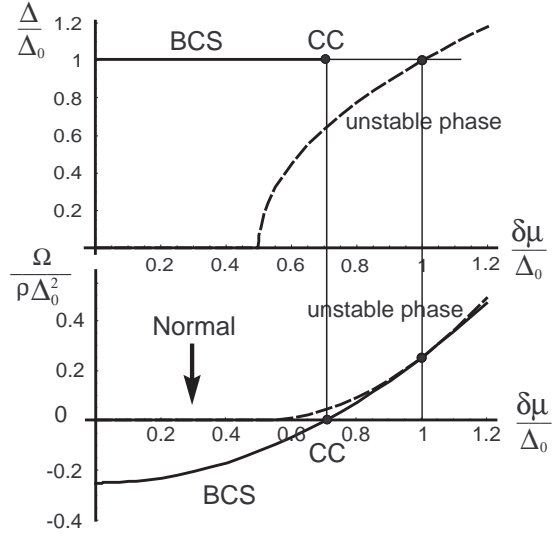}
  \caption{The two solutions of the gap equation with a mismatch
  $\delta\mu$. The continuous line is the BCS solution, the dashed
  one is called the Sarma solution.
  }\label{fig:1}
  \end{center}
\end{figure}
\begin{figure}[h]
\begin{center}
  \includegraphics[width=.6\textwidth]{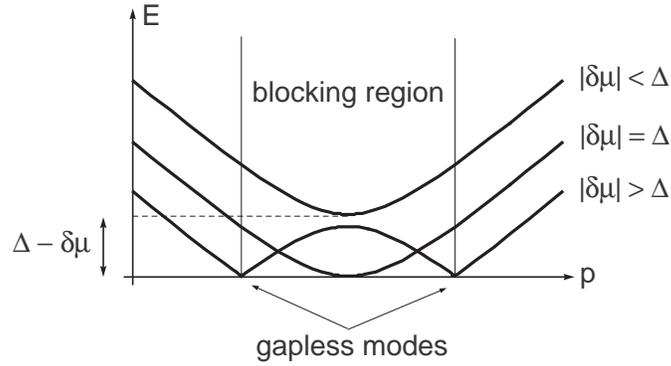}
 \caption{The spectrum of quasi-particles for different values of the mismatch $\delta\mu$.}
  \label{fig:2}
 \end{center}
\end{figure}
where $\Delta_0$ is the BCS solution of the gap equation for
$\delta\mu=0$ The free energy of the two solutions are given by \bea
a)&&~~~\Omega(\delta\mu)=\Omega_0(\delta\mu)-\dd\frac \rho
  4(-2\delta\mu^2+\Delta_0^2),\nn\\ b)&&~~~\Omega(\delta\mu)=\Omega_0(\delta\mu)
  -\dd\frac \rho
  4(-4\delta\mu^2+4\delta\mu\Delta_0-\Delta_0^2),\eea
with $\Omega_0(\delta\mu)$ the free energy for unpaired fermions.
For two massless fermions $p_F=\mu$ and $v_F=1$ and
$\rho={\mu^2}/{\pi}$.  The two solutions are illustrated in Fig.
\ref{fig:1}. We see that the solution a) is always favored with
respect to the solution b) (called the Sarma phase
\cite{sarma:1963sa}). Furthermore the BCS phase goes to the normal
phase at \be\delta\mu_1=\frac{\Delta_0}{\sqrt{2}}.\ee This point is
called the Chandrasekhar-Clogston (CC) point \cite{Chandrasekhar}
(denoted by CC in Fig. \ref{fig:1}). Ignoring for the moment that in
this case, after the CC point the system goes to the normal phase,
we notice that the gaps of the two solutions coincide at
$\delta\mu=\Delta_0$. This is a special point, since in presence of
a mismatch the spectrum of the quasi-particles is modified as
follows\be E_{\delta\mu=0}=\sqrt{(p-\mu)^2+\Delta^2}\to
E_{\delta\mu}=\left|\delta\mu\pm\sqrt{(p-\mu)^2+\Delta^2}\,\right|.\ee
Therefore for $|\delta\mu|<\Delta$ we have gapped quasi-particles
with gaps $\Delta\pm\delta\mu$ (see Fig. \ref{fig:2}). However, for
$|\delta\mu|=\Delta$ a gapless mode appears and from this point on
there are regions of the phase space which do not contribute to the
gap equation (blocking regions). The gapless modes are characterized
by \be E(p)=0\Rightarrow p=\mu\pm\sqrt{\delta\mu^2-\Delta^2}.\ee
Since the energy cost for pairing two fermions belonging to Fermi
spheres with mismatch $\delta\mu$ is $2\delta\mu$ and the energy
gained in pairing is $2\Delta$, we see that the fermions begin to
unpair for \be 2\delta\mu\ge 2\Delta.\ee These considerations will
be relevant for the study of the gapless phases when neutrality is
required.

\section{The g2SC Phase}
\begin{figure}[h]
\begin{center}
  \includegraphics[width=.7\textwidth]{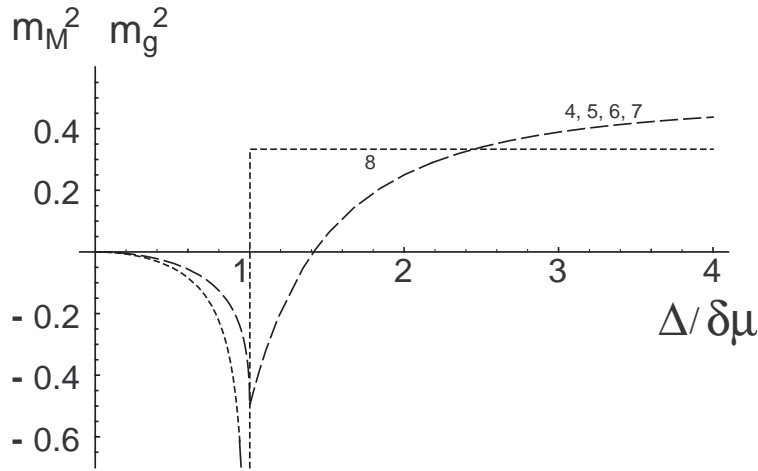}
  \caption{Plot of $m_M^2/m_g^2$ vs. $\Delta/\delta\mu$.
  Here $m_g^2=\mu^2  g^2/(3\pi^2)$.
  The long-dashed line corresponds to the gluons 4,5,6,7,
  whereas the short-dashed one to the gluon 8.}\label{fig:4}
  \end{center}
\end{figure}

The g2SC phase \cite{huang:2003ab} has the same condensate as the
2SC \be\langle
0|\psi_{aL}^{\alpha}\psi_{bL}^\beta|0\rangle=\Delta\epsilon^{\alpha\beta
3}_{ab 3},~~~\alpha,\beta\in SU_c(3),~~~a,b\in SU(2)_L,\ee and
technically, it is distinguished by  2SC due to  the presence of
gapless modes starting at $\delta\mu=\Delta$. In this case only two
massless flavors are present (quarks $u$ and $d$) and there are 2
quarks ungapped $q_{ub}, q_{db}$ and 4  gapped $q_{ur}$, $q_{ug}$,
$q_{dr}$, $q_{dg}$, where the color indices
 $1,2,3$ have been identified with $r,g,b$ (red, green and blue).
 The difference with the usual 2SC phase is that  color
 and electrical neutrality are required:\be \frac{\de \Omega}{\de\mu_e}=\frac{\de
\Omega}{\de\mu_3}=\frac{\de \Omega}{\de\mu_8}=0.\ee This creates a
mismatch between the two Fermi spheres given by \be
\delta\mu=\frac{p_F^d-p_F^u}2=\frac{\mu_d-\mu_u}2=\frac{\mu_e}2.\ee
Furthermore the gap equation must be satisfied \be \frac{\de
\Omega}{\de\Delta}=0.\ee
\begin{figure}[h]
\begin{center}
  \includegraphics[width=.6\textwidth]{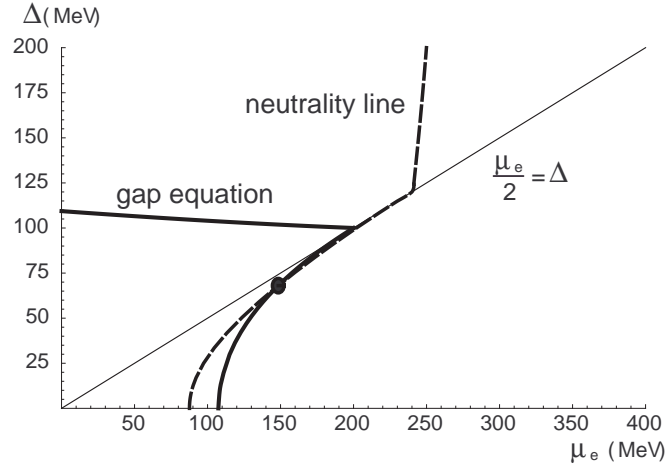}
  \caption{The plane $(\mu_e,\Delta)$ showing the lines of the solutions
  of the gap equation (continuous) and to the neutrality condition (dashed). The common solution
  is marked by a black dot.
  }\label{fig:3}
  \end{center}
\end{figure}
The solutions to these equations are plotted
in the plane $(\mu_e,\Delta)$ in Fig. \ref{fig:3}. In this figure we
see the two branches of solutions of the gap equation corresponding
to the BCS phase and to the Sarma phase (compare with Fig.
\ref{fig:1}). Therefore the solution to the present problem belongs
to the Sarma branch. In \cite{huang:2003ab} it is also shown that
the solution is a minimum of the free energy following the
neutrality line. On the other hand this point is a maximum following
the appropriate line $\mu_e=\rm{const.}$. We see that  the
neutrality conditions promote the unstable phase (Sarma) to a stable
one. However this phase has an instability connected to the Meissner
mass of the gluons \cite{huang_instability1}. In this phase the
color group $SU_c(3)$ is spontaneously broken to $SU_c(2)$ with 5 of
the 8 gluons acquiring a mass; precisely the gluons 4,5,6,7,8. At
the point $\delta\mu=\Delta$ where the 2SC phase goes into the g2SC
one, all the massive gluons have imaginary mass. Furthermore the
gluons 4,5,6,7 have imaginary mass already starting at
$\delta\mu=\Delta/\sqrt{2}$, that is at the Chandrasekhar-Clogston
point, see Fig. \ref{fig:4}. This shows that both  the g2SC and the
2SC phases are unstable. The instability of the g2SC phase seems to
be a general feature of the phases with gapless modes
\cite{alford_wang}.

%\end{document}
\section{The gCFL phase}

\begin{figure}[h]
\begin{center}
  \includegraphics[width=.6\textwidth]{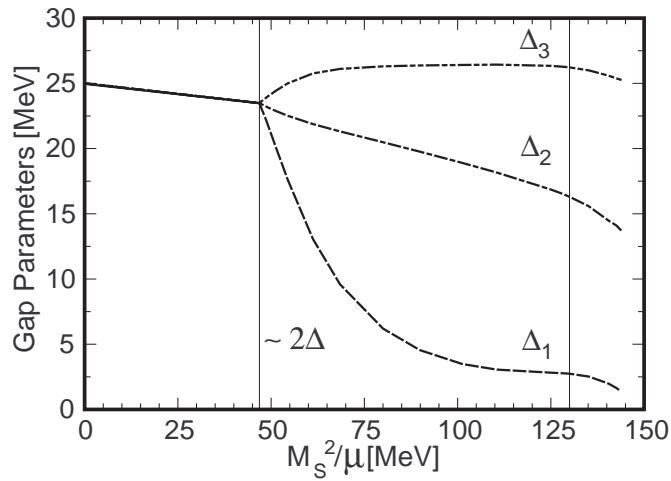}
  \caption{The behavior of the gap parameters in gCFL. The
  parameters has been chosen  in such a way that
  $\Delta_{0}=25~MeV$ and $\mu=500~MeV$ \cite{Alford:2004hz}. The vertical line at
  $M_s^2/\mu\approx 130 ~MeV$ marks the transition from the gCFL
  phase to the normal one.
  }\label{fig:5}
  \end{center}
\end{figure}

The gCFL phase is a generalization of the CFL phase which has been
studied both at $T=0$ \cite{Alford:2003fq,Alford:2004hz} and
$T\not=0$ \cite{gCFL_2}. The condensate has now the following form
\be \langle
0|\psi_{aL}^\alpha\psi_{bL}^\beta|0\rangle=\Delta_1\epsilon^{\alpha\beta
1}\epsilon_{ab1}+\Delta_2\epsilon^{\alpha\beta
2}\epsilon_{ab2}+\Delta_3\epsilon^{\alpha\beta 3}\epsilon_{ab3}.\ee
The CFL phase corresponds to all the three gaps $\Delta_i$ being
equal. Varying the gaps one gets many different phases. In
particular we will be interested to CFL, to g2SC characterized by
$\Delta_3\not=0$ and $\Delta_1=\Delta_2=0$ and to the gCFL phase
with $\Delta_3>\Delta_2>\Delta_1$. Notice that,  in the actual
context,  the strange quark is present also in the g2SC phase but
unpaired. The matrix of the condensates in the color ($r,g,b$) and
flavor ($u,d,s$) space is given below:\be
\begin{array}{|c|ccccccccc|}  \hline
 & ru & gd & bs & rd & gu & rs & bu & gs & bd \\ \hline
ru &  & \Delta_3 & \Delta_2 &  &  &  &  &  &  \\
gd & \Delta_3 &  & \Delta_1 &  &  &  &  &  &  \\
bs & \Delta_2 & \Delta_1 &  &  & &  &  &  &  \\
rd &  &  &  &  & -\Delta_3 &  &  &  &  \\
gu &  &  &  & -\Delta_3 &  &  &  &  &  \\
rs &  &  &  &  &  &  & -\Delta_2 &  &  \\
bu &  &  &  &  &  & -\Delta_2 &  &  &  \\
gs &  &  &  &  &  &  &  &  & -\Delta_1 \\
bd &  &  &  &  &  &  &  & -\Delta_1 &  \\  \hline
\end{array}
\ee In flavor space the gaps $\Delta_i$ correspond to the following
pairings \be \Delta_1\Rightarrow ds,~~~\Delta_2\Rightarrow us,~~~
\Delta_3\Rightarrow ud.\ee The mass of the strange quark is taken
into account by shifting all the chemical potentials involving the
strange quark as follows: \be\mu_{\alpha s}\to \mu_{\alpha s}
-\frac{M_s^2}{2\mu}.\ee It has also been shown in ref. \cite{alford}
that color and electric neutrality in CFL require \be
\mu_8=-\frac{M_s^2}{2\mu},~~~\mu_e=\mu_3=0.\ee At the same time the
various mismatches are given by \be
\delta\mu_{bd-gs}=\frac{M_s^2}{2\mu},~~~\delta\mu_{rd-gu}=\mu_e=0,~~~
\delta\mu_{rs-bu}=\mu_e-\frac{M_s^2}{2\mu}.\ee It turns out that in
the gCFL the electron density is different from zero and, as a
consequence, the mismatch between the quarks $d$ and $s$ is the
first one to give rise to the unpairing of the corresponing quarks.
This unpairing is expected to occur for \be
2\frac{M_s^2}{2\mu}>2\Delta~~
\Rightarrow~~\frac{M_s^2}{\mu}>2\Delta.\ee This has been
substantiated by the calculations in a NJL model modeled on one
gluon-exchange in \cite{Alford:2004hz}. The results for the gaps are
given in Fig. \ref{fig:5}. We see that the transition from the CFL
phase, where all gaps are equal, to the gapless phase occurs roughly
at $M_s^2/\mu =2\Delta$.
\begin{figure}[h]
\begin{center}
  \includegraphics[width=.6\textwidth]{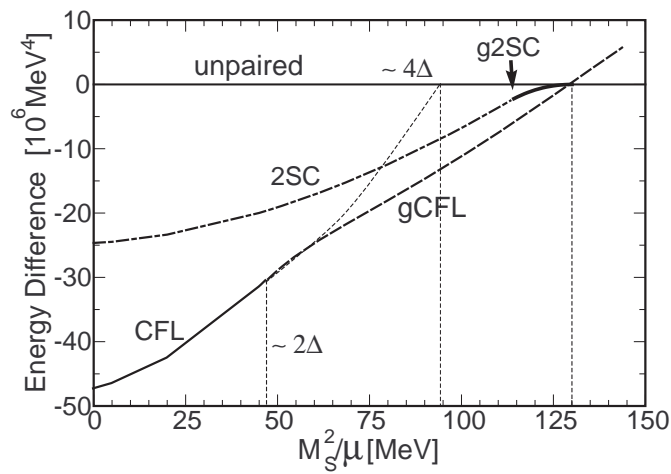}
  \caption{ We give here the free energy of the various phases with
  reference to the normal phase \cite{Alford:2004hz}, named unpaired in the figure.
  \label{fig:6}}
  \end{center}
\end{figure}
In Fig. \ref{fig:6} we show the free energy of the various phases
with reference to the normal phase. The CFL phase is the stable one
up to $M_s^2/\mu\approx  2\Delta$. Then the gCFL phase takes over up
to about 130 $MeV$ where the system goes to the normal phase. Notice
that except in a very tiny region around this point, the CFL and
gCFL phases win over the corresponding 2SC and g2SC ones. The thin
short-dashed line represents the free energy of the CFL phase up to
the point where it becomes equal to the free-energy of the normal
phase. This happens for $M_s^2/\mu\approx 4\Delta$. This point is
the analogue  of the Chandrasekhar-Clogston point of the two-flavor
case.

\begin{figure}[h]
\begin{center}
  \includegraphics[width=1\textwidth]{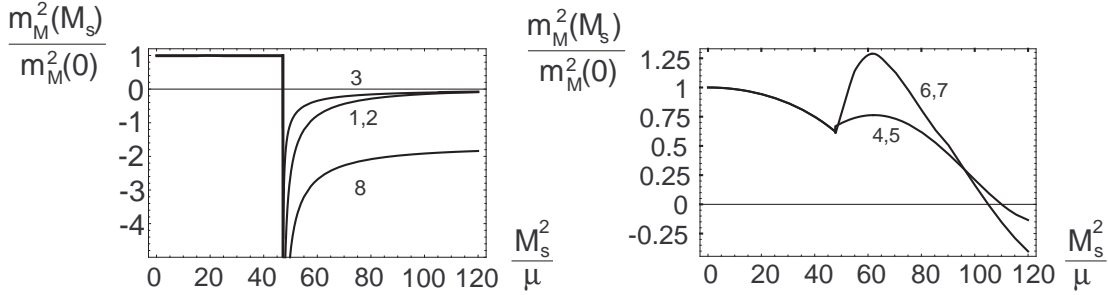}
  \caption{The figure shows, for the gCFL case, the masses of the gluons
  1,2,3, 8 (left panel) and 4,5,6,7 (right panel) vs. $M_s^2/\mu$.
  \label{fig:7}}
  \end{center}
\end{figure}

The gCFL phase has gapless excitations and, as a consequence, the
chromomagnetic instability discussed in the case of the g2SC phase
shows up  here too. This has been shown in
\cite{MeissnerCFL_1,MeissnerCFL_2}. The results of ref.
\cite{MeissnerCFL_1} are given in Fig. \ref{fig:7} for the various
gluon masses.

The existence of the chromomagnetic instability is a serious problem
for the gapless phases (g2SC and gCFL) but also for the 2SC phase,
as we have discussed previously. A way out of this problem would be
to have gluon condensation. For instance, if one assumes
artificially $\langle A_\mu^3\rangle$ and $\langle A_\mu^8\rangle$
not zero and with a value of about 10 $MeV$ it can be shown that the
instability disappears \cite{MeissnerCFL_1}. Also, very recently in
\cite{miransky}, it has been shown the possibility of eliminating
the chromomagnetic instability in the 2SC phase through a gluonic
phase. However it is not clear if the same method can be extended to
the gapless phases.

Another interesting possibility has been considered in three papers
by Giannakis and Ren, who have considered the LOFF phase, that is a
nonhomogeneous phase first studied in a condensed matter context
\cite{LOFF1,LOFF2} and then in QCD in
\cite{Alford:2000ze,Bowers:2002xr} (for recent reviews of the LOFF
phase, see \cite{Casalbuoni:2003wh,Bowers:2003ye}). The results
obtained by Giannakis and Ren in the two-flavor case are the
following:
\begin{itemize}
  \item The presence of the chromomagnetic instability in g2SC is
  exactly what one needs in order that the LOFF phase is
  energetically favored \cite{Ren1}.
  \item  The LOFF phase in the two-flavor case has no chromomagnetic
  instabilities (though it has gapless modes) at least in the weak
  coupling limit \cite{Ren2,Ren3}.
\end{itemize}
Of course these results make the LOFF phase a natural candidate for
the stable phase of QCD at moderate densities. In the next Sections
we will describe the LOFF phase in its simplest version and a very
simple approach to the problem with three flavors.

\section{The LOFF Phase}

\begin{figure}[ht]
\begin{center}
\includegraphics[width=.4\textwidth]{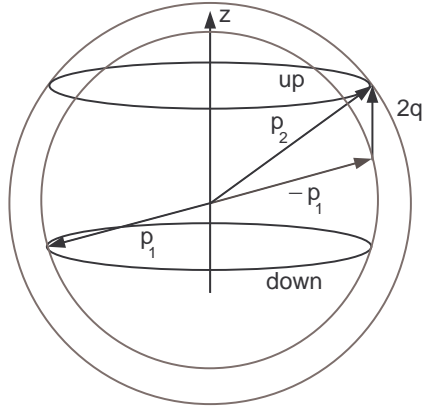}
\caption{Pairing of fermions belonging to two Fermi spheres of
different radii according to LOFF. \label{fig:8}} \end{center}
\end{figure}

According to the authors of refs. \cite{LOFF1,LOFF2} when fermions
belong to  different Fermi spheres, they  might prefer to pair
staying as much as possible close to their own Fermi surface. When
they are sitting exactly at the surface, the pairing is as shown in
Fig. \ref{fig:8}.
We see that the total momentum of the pair is ${\vec p}_1+{\vec
p}_2=2\vec q$ and, as we shall show, $|\vec q\,|$ is fixed
variationally whereas the direction of $\vec q$ is chosen
spontaneously. Since the total momentum of the pair is not zero the
condensate breaks rotational and translational invariance. The
simplest form of the condensate compatible with this breaking is
just a simple plane wave (more complicated possibilities will be
discussed later) \be \langle\psi(x)\psi(x)\rangle\approx\Delta\,
e^{2i\vec q\cdot\vec x}.\label{single-wave}\ee It should also be
noticed that the pairs use much less of the Fermi surface than they
do in the BCS case. In fact, in the case considered in Fig.
\ref{fig:8} the fermions can pair only if they belong to the circles
in  figure. More generally there is a quite large region in momentum
space (the so called blocking region) which is excluded from
pairing. This leads to a  condensate generally smaller than the BCS
one.

Let us now consider in more detail the LOFF phase. For two fermions
at different densities  we have an extra term in the hamiltonian
which can be written as \be
H_I=-\delta\mu\sigma_3\label{interaction},\ee where, in the original
LOFF papers \cite{LOFF1,LOFF2} $\delta\mu$ is proportional to the
magnetic field due to the impurities, whereas in the actual case
$\delta\mu=(\mu_1-\mu_2)/2$ and $\sigma_3$ is a Pauli matrix acting
on the two fermion space. According to refs. \cite{LOFF1,LOFF2} this
favors the formation of pairs with momenta $\vec p_1=\vec k+\vec
q,~~~\vec p_2=-\vec k+\vec q$. We will discuss in detail the case of
a single plane wave (see eq. (\ref{single-wave})). The interaction
term of eq. (\ref{interaction}) gives rise to a shift in $\xi$ (see
eq. (\ref{xi})) due both to the non-zero momentum of the pair and to
the different chemical potentials \be \xi=E(\vec p)-\mu\to E(\pm\vec
k+\vec q)-\mu\mp\delta\mu\approx \xi\mp\bar\mu,\ee with \be
\bar\mu=\delta\mu-{\vec v}_F\cdot\vec q.\ee Notice that the previous
dispersion relations show the presence of gapless modes at momenta
depending on the angle with $\vec q$. Here we have assumed
$\delta\mu\ll\mu$ (with $\mu=(\mu_1+\mu_2)/2$) allowing us to expand
$E$ at the first order in $\vec q/\mu$ (see Fig. \ref{fig:8}).

\begin{figure}[ht]
\begin{center}
\includegraphics[width=.9\textwidth]{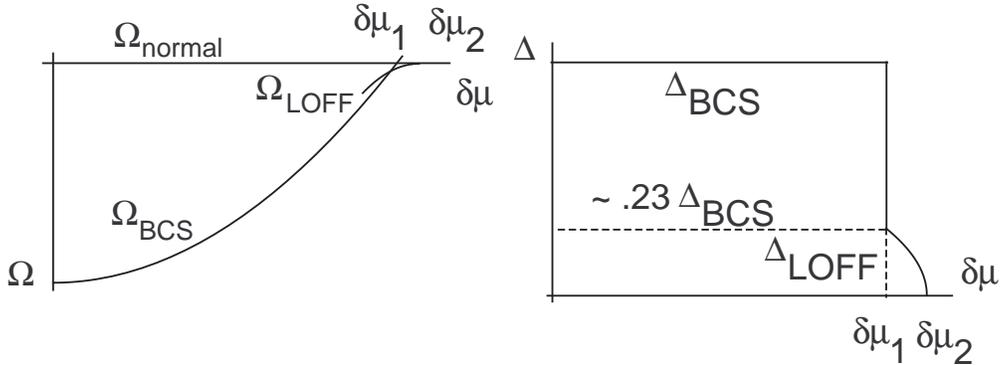}
\caption{The grand potential (left panel) and the condensates of the
BCS and LOFF phases vs. $\delta\mu$ (right panel). \label{fig:9}}
\end{center}
\end{figure}

The gap equation for the present case is obtained simply from eq.
(\ref{gap}) via the substitution \be\delta\mu\to\bar\mu.\ee By
studying eq. (\ref{gap}) one can  show that increasing $\delta\mu$
starting from zero, we have first the BCS phase. Then  at
$\delta\mu=\delta\mu_1$ there is a first order transition to the
LOFF phase \cite{LOFF1,Alford:2000ze}, and at
$\delta\mu=\delta\mu_2>\delta\mu_1$ there is a second order phase
transition to the normal phase \cite{LOFF1,Alford:2000ze}. We start
comparing the grand potential in the BCS phase to the one in the
normal phase. Their difference is given by \be \Omega_{\rm
BCS}-\Omega_{\rm
normal}=-\frac{p_F^2}{4\pi^2v_F}\left(\Delta^2_0-2\delta\mu^2\right),\ee
where the first term comes from the energy necessary to the BCS
condensation, whereas the last term arises from the grand potential
of two free fermions with different chemical potential. We recall
also that for massless fermions $p_F=\mu$ and $v_F=1$. We have again
assumed $\delta\mu\ll\mu$. This implies that there should be a first
order phase transition from the BCS to the normal phase at
$\delta\mu=\Delta_0/\sqrt{2}$ \cite{Chandrasekhar}, since the BCS
gap does not depend on $\delta\mu$. The situation is represented in
Fig. \ref{fig:9}.
In order to compare with the LOFF phase one can  expand the gap
equation around the point $\Delta=0$ (Ginzburg-Landau expansion)  to
explore the possibility of a second order phase transition
\cite{LOFF1}. The result for the free energy is \be \Omega_{\rm
LOFF}-\Omega_{\rm normal}\approx
-0.44\,\rho(\delta\mu-\delta\mu_2)^2.\ee At the same time, looking
at the minimum in $q$ of the free energy one finds \be qv_F\approx
1.2\, \delta\mu.\label{q}\ee

 We see that in the window
between the intersection of the BCS curve and the LOFF curve in Fig.
\ref{fig:9} and $\delta\mu_2$, the LOFF phase is favored. Also at
the intersection there is a first order transition between the LOFF
and the BCS phase. Furthermore, since $\delta\mu_2$ is very close to
$\delta\mu_1$ the intersection point is practically given by
$\delta\mu_1$. In Fig. \ref{fig:9} we show, in the right panel, the
behaviour of the condensates. Although the window
$(\delta\mu_1,\delta\mu_2)\simeq(0.707,0.754)\Delta_0$ is rather
narrow, there are indications that, considering the realistic case
of QCD \cite{Leibovich:2001xr}, the window opens up. Such opening
occurs also for different crystalline structures than the single
plane wave considered here \cite{Bowers:2002xr,Casalbuoni:2004wm}.

\section{The LOFF phase with three flavors}

\begin{figure}[h]
\begin{center}
\includegraphics[width=.5\textwidth]{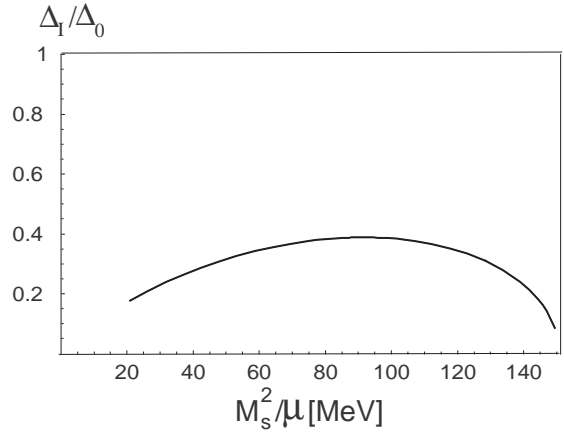}
\caption{The gap for LOFF with three flavors vs. $M_s^2/\mu$. The
line corresponds to the most favored solution, that is to the
configurations 1 and 2. \label{fig:11}}
\end{center}
\end{figure}

\begin{figure}[h]
\begin{center}
\includegraphics[width=.5\textwidth]{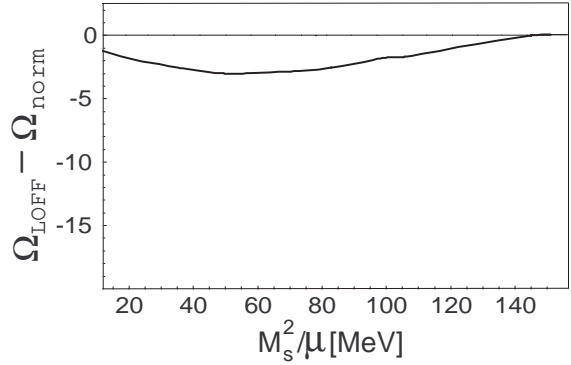}
\caption{The free energy of the most favored  configurations (1 and
2) considered for LOFF with three flavors vs.
$M_s^2/\mu$.\label{fig:12}}\end{center}
\end{figure}

In the last Section we would like to illustrate some preliminary
result about the LOFF phase with three flavors. This problem has
been considered in \cite{casalbuoni_loff3} under various simplifying
hypothesis:
\begin{itemize}
  \item The study has been made in the Ginzburg-Landau
  approximation.
  \item Only electrical neutrality has been required and the
  chemical potentials for the color charges $T_3$ and $T_8$ have
  been put equal to zero (see later).
  \item The mass of the strange quark has been introduced as it was
  done previously previously for the gCFL phase.
  \item The study has been restricted to plane waves, assuming the
  following generalization of the gCFL case:
  \be \langle\psi^\alpha_{aL}\psi^\beta_{bL}\rangle=\sum_{I=1}^3\Delta_I(\vec x)
\epsilon^{\alpha\beta I}\epsilon_{ab I},~~~\Delta_I(\vec x)=\Delta_I
e^{2i\vec q_I\cdot\vec x}\ee
\item The condensate depends on three momenta, meaning three lengths
of the momenta $q_i$ and three angles. In \cite{casalbuoni_loff3}
only four particular geometries  have been considered: 1) all the
momenta parallel pointing upward the $z$-axis, then 2), 3) and 4)
are obtained by inverting respectively the momentum $\vec q_1$,
$\vec q_2$ and $\vec q_3$.
%\begin{figure}[ht]
%\begin{center}
%\includegraphics[width=.5\textwidth]{fig10.eps}
%\caption{The four configurations of the vectors $q_i$ considered in
%the study of LOFF with three flavors.}\label{fig:10}\end{center}
%\end{figure}

\end{itemize}
Under the previous hypothesis the free energy (with reference to the
normal state) has the expansion \be \Omega -\Omega_{normal}=
\sum_{I=1}^3\left(\frac{\alpha_I}{2}\,\Delta_I^2 ~+~
\frac{\beta_I}{4}\,\Delta_I^4 ~+~ \sum_{I\neq
J}\frac{\beta_{IJ}}{4}\,\Delta_I^2\Delta_J^2 \right) ~+~
O(\Delta^6)\ee with \be\alpha_I(q_I,\delta\mu_I)=-
\frac{4\mu^2}{\pi^2} \left(1 -
\frac{\delta\mu_I}{2q_I}\log\left|\frac{q_I+\delta\mu_I}{q_I-\delta\mu_I}\right|
- \frac{1}{2}\log\left|\frac{4(q_I^2
-\delta\mu_I^2)}{\Delta_0^2}\right|\right)\ee \be
\beta_I(q_I,\delta\mu_I)=\frac{\mu^2}{\pi^2}\frac{1}{q_I^2-\delta\mu_I^2}\ee
\be \beta_{12}=-2\frac{\mu^2}{\pi^2} \int\frac{d{\bf
n}}{4\pi}\,\frac{1}{(2{\bf q_1}\cdot{\bf
n}+\mu_s-\mu_d-i0^+)\,(2{\bf q_2}\cdot{\bf n}+\mu_s-\mu_u-i0^+)}\ee
and the other $\beta_{IJ},~I\not = J$ obtained by the exchange \be
12\rightarrow 23,~ \mu_s\leftrightarrow\mu_d,~~~12\rightarrow 13,~
\mu_s\leftrightarrow\mu_u\ee The $\delta\mu_I$ are obtained from
\be\mu_u=\mu-\frac 2 3 \mu_e,~~\mu_d=\mu+\frac 1 3\mu_e,~~
\mu_s=\mu+\frac 1 3\mu_e-\frac{M_s^2}{2\mu}\label{eq:6.7}\ee In
particular the coefficients of $\Delta_I^2$ are the same as for LOFF
with two flavors. Therefore the minimization with respect to the
$|\vec q_I|$'s leads to the same result as in eq. (\ref{q}) \be
|\vec q_I|=1.2 \delta\mu_I\label{eq:55}\ee Then, one has to minimize
with respect to the gaps and to $\mu_e$ in order to require
electrical neutrality. It turns out that the configurations 3 and 4
have an extremely small gap. Furthermore for $M_s^2/\mu$ greater
than about 80 $MeV$ one has a solution with $\Delta_1=0$ and
$\Delta_2=\Delta_3$. In this case the configurations 1` and 2 have
the same free energy.  The results for the free energy and for the
gap of this solution are given in Fig. \ref{fig:11} and
\ref{fig:12}. In this study, the following choice of the parameters
has been made:  the BCS gap, $\Delta_0=25~MeV$, and the chemical
potential $\mu=500~MeV$. The values are the same discussed
previously for gCFL in order to allow for a comparison of the
results.

\begin{figure}[t]
\begin{center}
\includegraphics[width=.55\textwidth]{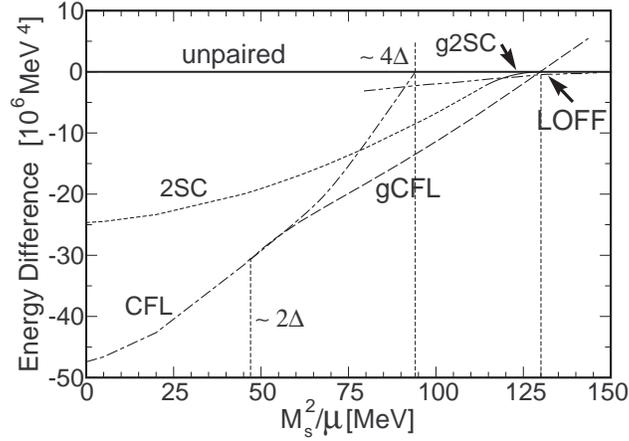}
\caption{Comparison of the free energy of the various phases with
the LOFF phase with three flavors.} \label{fig:13} \end{center}
\end{figure}

We are now in the position to compare these results with the ones
obtained in \cite{Alford:2004hz} for the gCFL phase. The comparison
is made in Fig. \ref{fig:13}. Ignoring the chromomagnetic
instabilities of the gapless phases and of 2SC we see that LOFF
takes over with respect to gCFL at about $M_s^2/\mu=128~MeV$ and
goes over to the normal phase for $M_s^2/\mu\approx 150~MeV$.

However, since  the instability exists it should be cured  in some
way. The results for the LOFF phase, assuming that also for three
flavors the chromomagnetic instability does not show up, say that it
could be the LOFF phase to takes over the CFL phase before the
transition to gCFL. For this it is necessary that the window for the
LOFF phase gets enlarged. However, in \cite{Casalbuoni:2004wm} it
has been show that for structures more general than the plane wave
the windows may indeed becomes larger. If define the window for the
single plane wave as $(\delta\mu_2-\delta\mu_1)/\delta\mu_2$ (see
the previous Section) we would get 0.06. The analogous ratio in
going from one to three plane waves goes to about
$(150-115)/150=.23$, with a gain of almost a factor 4. On the other
hand, in \cite{Casalbuoni:2004wm} it has been shown that considering
some of the crystalline structures already taken in exam in
\cite{Alford:2000ze}, as the face centered cube or the cube the
windows becomes about $(1.32-0.707)/1.32=0.46$ with a gain of about
7.7 with respect to the single plane wave. If these gains would be
maintained in going from two to three flavors with the face centered
cube structure, one could expect a gain from 4 to 7.7 with an
enlargement of the window between 88 and 170 $MeV$, which would be
enough to cover the region of gCFL (which is about 70 $MeV$).

\begin{figure}[h]\begin{center}
\includegraphics[width=.9\textwidth]{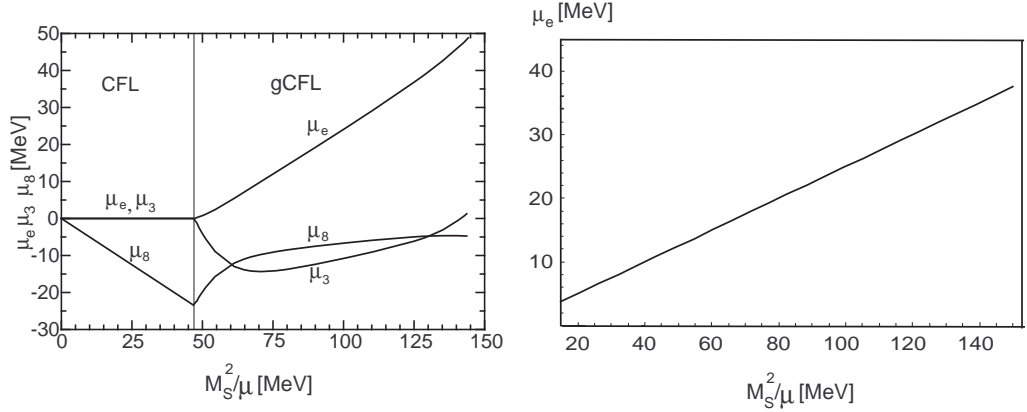}
\caption{The chemical potential for gCFL (left panel) and LOFF
(right panel) vs. $M_s^2/\mu$. \label{fig:14}}\end{center}
\end{figure}

At last we want to comment about the approximation in neglecting the
color neutrality condition and assuming $\mu_3=\mu_8=0$. In Fig.
\ref{fig:14} we show the chemical potentials
$\mu_e,~\mu_3,~\mu_8$for the gCFL phase in the left panel, and
$\mu_e$ for the LOFF phase in the right panel. We can make two
observations: first of all, in the region of interest where LOFF
dominates over gCFL the behaviour of $\mu_e$ in the two phases is
pretty much similar, and $\mu_3,~\mu_8\ll\mu_e$ for gCFL. This
suggests that also in the LOFF case $\mu_3$ and $\mu_8$ are small.
Second and more important the result of \cite{casalbuoni_loff3}
shows that $\mu_e\approx M_s^2/(4\mu)$ as for the case of 3 color
and 3 flavor unpaired quarks \cite{alford}. As can be seen from eq.
(\ref{eq:6.7}) this coreesponds to a symmetrical spli of the $s$ and
$d$ Fermi surfaces around the $u$ Fermi surface. Therefore  ${\bf
q}_2$ = ${\bf q}_3$ and the gaps $\Delta_2$ and $\Delta_3$ must
coincide. At the same time the separation of the $d$ and $s$
surfaces is the double and therefore $\Delta_1=0$. The unpaired
quarks have also $\mu_3=\mu_8=0$. Also, from Fig. \ref{fig:11} we
see that in our approximations the transition from the LOFF to the
normal phase is very close to be continuous. Since we expect also
the chemical potentials to be continuous at the transition point
very close to the critical point we should have $\mu_3=\mu_8=0$ also
on the LOFF side. This means the color neutrality condition should
be $\mu_3=\mu_8=0$ in the neighborhood of the transition. Therefore
we expect  the determination of the point $M_s^2/\mu=150~MeV$ to be
safe. On the other hand, the requirement of color neutrality could
change the intersection point with gCFL. Nevertheless, since the
critical point for LOFF is higher than the one of gCFL, for
increasing $M_s$ the system must to go into the LOFF phase.

%\newpage

\end{document}